# Advanced method for reliable estimation of the spin-orbit torque efficiency in low coercive ferromagnetic multilayers

M.E. Stebliy[1*], A.G. Kolesnikov[1], A.V. Ognev[1], E.V. Stebliy[1], X. Wang[2], H.F. Han[2], A.S. Samardak[1,3*]

[1]School of Natural Sciences, Far Eastern Federal University, Vladivostok 690950, Russia
[2]Beijing National Laboratory for Condensed Matter Physics, Institute of Physics, Chinese Academy of Sciences, Beijing 100190, China
[3]National Research South Ural State University, Chelyabinsk 454080, Russia
*email: steblii.me@dvfu.ru; samardak.as@dvfu.ru

An experimental study of current-induced magnetization reversal of the Ru/Co/Ru and Ru/Co/Ru/W structures was carried out. In the considered structures, due to the small value of the coercive force comparable in magnitude to the Oersted field and the SOT effect field, magnetization reversal is carried out by moving a domain wall parallel to the direction of current injection. For such a case, a new method for estimating the effective field of SOT based on the analysis of the domain wall position taking into account the distribution of the Oersted field was proposed. This method allowed determining the effective longitudinal field $H_L/j = 1.6 \cdot 10^{-11} mT/Am^{-2}$ and the efficiency of SOT $\xi_L = 0.03$ in the quasi-symmetric Ru/Co/Ru structure. It was found that adding the W capping layer enhances the SOT effect by 5 times.

## I. INTRODUCTION

The development of a reliable method for the local control of the magnetization is one of the key directions in the development of spintronics. A promising method is the current-induced switching of the magnetization in systems with a strong spin-orbit interaction due to the spin-orbit torque (SOT) [1]. This effect may be due to the spin Hall effect (SHE) and the Rashba effect (RE) and is usually realized in structures with perpendicular magnetic anisotropy (PMA) and broken inversion symmetry [2]. The simplest way to create such an asymmetry in order to realize deterministic magnetization switching is to apply an external in-plane magnetic field. In recent years, an active study of this mechanism has been aimed at searching for structures and optimizing process parameters, which make it possible to achieve maximum efficiency of current-induced magnetization reversal, as well as a theoretical explanation of the SOT effect mechanism.

To implement the SOT effect, three-layer structures consisting of a heavy metal (HM) buffer layer with a strong spin-orbit interaction, a ferromagnetic layer (FM) and a capping layer preserving oxidation and PMA are usually used. As a buffer, *5d* metals are mainly used (Pt, Ir, W, Ta, Hf) due to a combination of a large spin Hall effect and a small spin diffusion length ($\lambda_{sd}$) [3-5]. For this reason, structures containing 4d metals, for example, Ru or Pd, are much less studied [6, 7]. One of the ways to enhance the efficiency of current-induced magnetization reversal is the combination of various heavy metals, which can lead to an increase in the spin current density injected into the FM layer. Previously, structures with FM sandwiched between a pair of HM were studied: Pt-Ta [8], Pt-W [9]. As criteria for evaluating efficiency, the following parameters are usually chosen: the magnitude of longitudinal (antidamping) or transverse (field-like) fields ($\mu_0 H_{L/T}$) generated by the SOT effect [10, 11]; the magnitude of the critical magnetization reversal current ($I_c$) [12]; the effectiveness of the SOT effect ($\xi_j$) [13]. A general method for determining the magnitude of the effective magnetic field induced by the SOT effect is the harmonic approach based on calculation [13] or fitting [14]. There are other estimation methods, for example, the current-induced hysteresis loop shift method [15] or the approach based on the measurement of magnetization curves [16].

In this paper, an alternative method of direct estimation of the SOT field in low-coercive structures by the position of a domain wall is proposed. In addition to SHE and RE, as a result of the passing of current through a Hall bar, the Oersted field is generated, which is responsible for initiating the reversal process [17, 18]. Accounting for this field is especially important when considering systems with low coercive force.

## II. EXPERIMENTAL PROCEDURES

In a previous study, we showed that the magnetic parameters of Co, such as the coercive force and the PMA energy, in quasi-symmetrical Ru/Co/Ru films can vary over a wide range of values by changing the thickness of the FM and HM layers [19]. Such a possibility could make this system interesting for studying the dependence of the current-induced magnetization reversal effect on the magnitude of the magnetic parameters. However, experimental studies have shown that the effectiveness of SOT in such a system is small, as is shown below. To increase the efficiency in the considered structure, a capping layer of W was added. On the one hand, W has a large SHE angle ($\Theta_{SH}$) compared to Ru: -0.18 and +0.006 for W and Ru, respectively [6, 20]. On the other hand, W has the opposite to Ru sign of $\Theta_{SH}$ when these materials are placed from different sides of a ferromagnet. It leads to the summation of the spin currents from the bottom and top interfaces, Fig.1(a). Our experimental studies have shown that the presence of an ultrathin Ru interlayer between Co and W is a necessary condition for the preservation of PMA. The minimal possible thickness of the Ru interlayer, at which PMA is still induced, is 1.2 nm. We suggest that a Ru interlayer of this thickness should not be a source of spin polarized current, since its thickness is less than the spin diffusion length in Ru — 4 nm [21, 22].

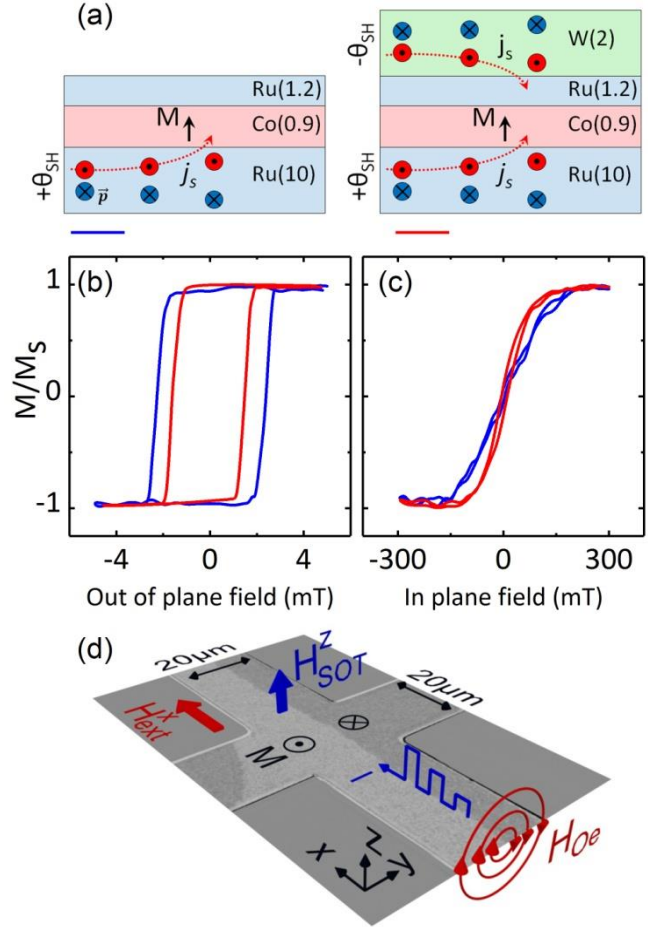

**FIG.1.** (a) Schemes of experimentally studied multilayer structures (thickness in nm). The red arrows indicate the direction of the spin current injection from Ru and W layers. Magnetic hysteresis loops measured with VSM during remagnetization in an external magnetic field applied perpendicular (b) and parallel (c) the sample plane. (d) Kerr image of the Hall bar under study with a schematic indication of the relative orientation of the magnetization **M**, the external field ($\mu_0 H_{ext}^x$), passing current ($I$), the Oersted field ($\mu_0 H_{Oe}$) induced by this current and the effective field of SOT ($\mu_0 H_{SOT}^z$)

At nanometer thicknesses, the W layer can be in amorphous-like, α-(bcc) or β-(A15 cubic) phase, for which the value of $\Theta_{SH}$ is very different: 0.23, 0.07 and 0.4, respectively [20, 23, 24]. Depending on the structure and parameters of the experiment, β-phase can form in the W thickness range up to 26 nm [25], but it has been defined experimentally that the largest $\Theta_{SH}$ value usually lies in the interval from 2.5 to 9 nm [9, 24]. The quantitative criterion for estimating the type of phase can be the value of resistivity: for α-phase it is from 20 to 40 μΩ·cm, for β-phase - from 120 to 260 μΩ·cm [9, 20, 25]. Intermediate resistivity values correspond to the metastable β-phase [26] or the mixed α+β phase, for which $\Theta_{SH}$ can take an intermediate value of ~0.18 [20]. We determined the phase of W by the resistivity measurement.

We measured the specific resistance of the W capping layer for the series of Ru(10)/Co(0.9)/Ru(1.2)/W($t_W$) samples, where $t_W$ changes from 0 to 8 nm, Fig.2. Comparing the obtained values of specific resistance with the literature data, we can conclude that the W layer is in the mixed α+β phase. It was determined experimentally that a sharp increase in the efficiency of current-induced magnetization reversal, as compared with the Ru/Co/Ru structure, is observed up to $t_W$=2 nm. A further increase in the W thickness leads to a slight linear enhancement in the magnitude of the effective fields induced by the current. Thus, the sharp increase in the SOT efficiency is completed when $t_W$ is comparable to the spin diffusion length, which for W lies in the range from 1.1 to 3.5 nm [24, 27].

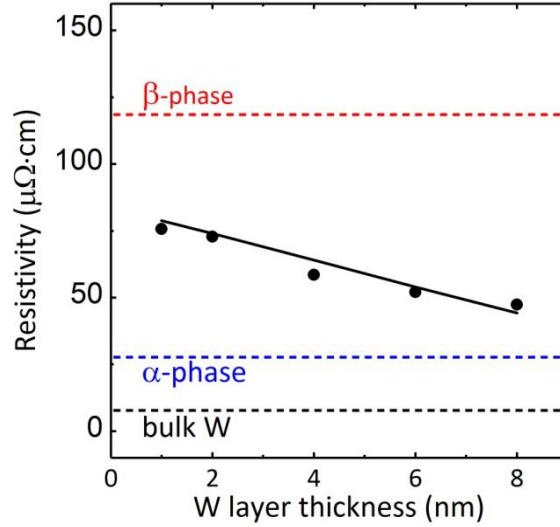

**FIG.2**. Experimental values of the specific resistance of the W capping layer in Ru(10)/Co(0.9)/Ru(1.2)/W(tw) structures. The dotted line in the graph shows the average values of the specific resistance corresponding to α and β phases, as well as to the bulk W taken from the literature data.

Experimental samples based on Ru/Co/Ru films were prepared by magnetron sputtering on an oxidized SiO$_2$ silicon substrate. The base pressure in the vacuum system was $1.3 \times 10^{-6}$ Pa, DC sputtering of all materials was carried out in the Ar atmosphere of pressure of 0.5 Pa at room temperature. The deposition rates of materials were for Co - 0.07 nm/s; Ru - 0.02 nm/s; W - 0.05 nm/s. For the study of current-induced magnetization reversal, Hall bars with a size of 20×200 μm$^2$ were prepared using the photolithography and lift-off procedure, Fig.1d. The magnetic parameters of the structures were measured using a vibration sample magnetometer (VSM 7410, LakeShore). Current-induced magnetization reversal was investigated at a probe station equipped with electromagnets inducing in-plane ($\mu_0 H_{ext}^x$) and out-of-plane ($\mu_0 H_{ext}^z$) fields, together with a Keithley 2182a voltmeter and a Keithley2420 current source. The current-induced magnetization reversal process was visualized using a Kerr microscope.

In this paper, we study the effect of current on the magnetization in two structures: Ru(10)/Co(0.9)/Ru(1.2) и Ru(10)/Co(0.9)/Ru(1.2)/W(2), the thickness is in nanometers. The magnetic parameters of the prepared samples, measured for the corresponding films, are given in Table 1. Comparing the results, it can be noted that with the addition of the W layer, the magnitude of the saturation magnetization $M_s$ do not change, whereas the value of the coercive force $\mu_0 H_c$ and the energy of effective anisotropy $K_{eff}$ (or the field $\mu_0 H_k^{eff}$) change in two or more times. In Ru/Co/Ru films, the magnetoelastic anisotropy makes the main contribution to PMA [19]. It can be assumed that W atoms penetrate into the Ru layer due to their high diffusion capacity and form impurity dislocations that reduce the stresses in the Ru/Co interface, the effective PMA and coercive force [28, 29].

**Table I**

Magnetic parameters of *Ru(10)/Co(0.9)/Ru(1.2)* and *Ru(10)/Co(0.9)/Ru(2)* films

|  | Ru/Co/Ru | Ru/Co/Ru/W |
|---|---|---|
| $M_s, A/m$ | $1.1 \times 10^6$ | $1.1 \times 10^6$ |
| $\mu_0 H_k^{eff}, mT$ | 140 | 60 |
| $K_{eff}, J/m^3$ | $7.7 \times 10^4$ | $3.3 \times 10^4$ |
| $\mu_0 H_c, mT$ | 2.2 | 1.6 |

### III. RESULTS AND DISCUSSION
#### 3.1. Current-induced magnetization reversal

To study current-induced magnetization reversal, current pulses were passed through the Hall bars with a maximum amplitude of ± 60 mA and a duration of 7 ms in the presence of a planar field $H_{ext}^x = -8, 0, +8 \, mT$. The value of the field 8 mT was chosen for the reason that for the structure containing W in this field a complete switching of the magnetization begins to be observed. Passing the current led to the generation of two fields: the Oersted field $\mu_0 H_{Oe}$,, which has a non-uniform profile and a different orientation at opposite edges; and the effective field of SOT effect $\mu_0 H_{SOT}^z$, which is uniform across the entire microstrip, Fig.1(d). The change in magnetization was recorded using Kerr microscopy or by measuring the anomalous Hall effect (AHE), while passing the current of 1 mA. The potential difference resulting from AHE is proportional to the perpendicular component of the magnetization.

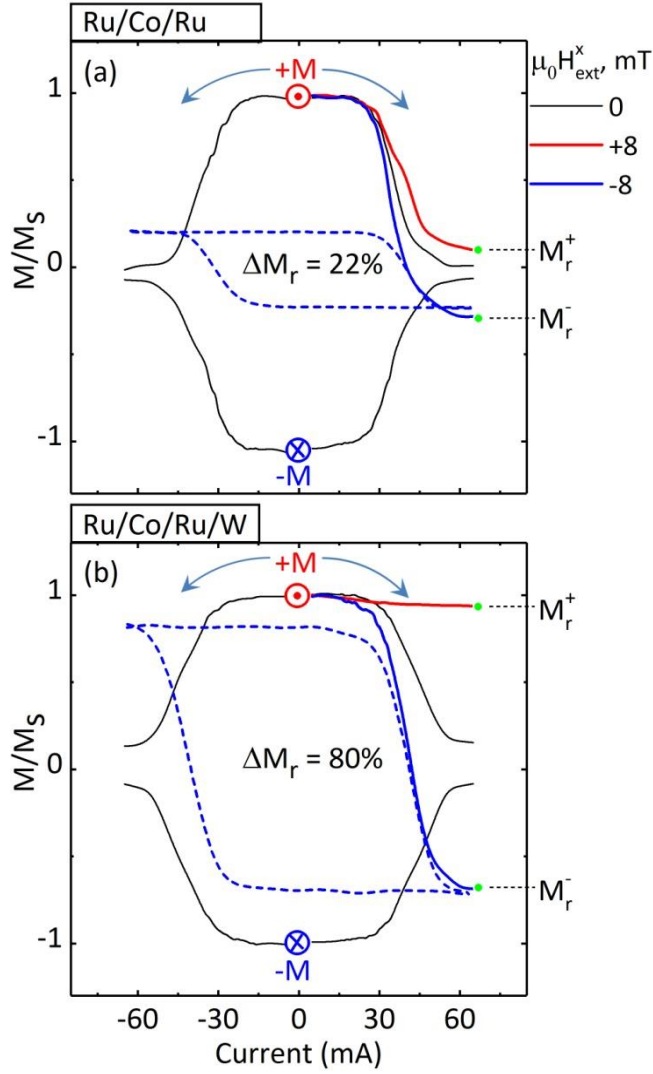

**FIG.3**. A set of demagnetization curves obtained by passing the current of increasing amplitude after preliminary magnetization to the saturation state with a different value of the in-plane field $\mu_0 H_{ext}^x$ for the Ru/Co/Ru (a) and Ru/Co/Ru/W (b). Current-induced magnetic hysteresis loops allow determining of the switching amplitude

For the convenience of visualization and further analysis, the effect of the current on a sample preliminarily magnetized to the saturation state was considered. We conducted the following sequence of measurements:

1. *Current-induced magnetization reversal in the absence of an external field $\mu_0 H_{ext}^x$*. Initially, the samples were magnetized by an external magnetic field $\mu_0 H_{ext}^z$ to a saturation state with $\pm M_s$. After that, in the absence of an external magnetic field, current pulses with amplitude from 0 to $\pm 60$ mA with a step of 2 mA were applied. After each pulse the AHE value was measured. The current density was not greater than $j_{max} \approx 2 \cdot 10^{11} A/m^2$. The measured value of AHE is proportional to the change in the sample magnetization (M). The maximum value of AHE corresponds to the saturation state with $\pm M_s$, and it is minimal when M → 0, i.e. at a demagnetized state. For a better presentation of the data, we used reduced $M/M_s$ values proportional to AHE. The results of measuring AHE in $\mu_0 H_{ext}^x = 0$ are indicated by the black line in Fig.3. It can be seen that with increasing current amplitude, the sample magnetization $\pm M_s$ decreases sharply at the current of $\pm 30$ mA and reaches saturation when I = $\pm 55$ mA and with M → 0.

The measurements were repeated for four combinations of the orientation of the initial magnetization $\pm M_s$ and the current direction $\pm I$. In all cases, the magnitude of the residual magnetization $M_r$ is the same in magnitude, which indicates the homogeneity of the system and the absence of effective magnetic fields $\mu_0 H_{SOT}^z$.

The results of visualization of the magnetization reversal process by the current for both structures are shown in Fig.4. It should be noted several differences between the conclusions on the magnetization reversal process defined from the magnetization reversal curves and the corresponding Kerr microscopy images. As it can be seen from the Kerr images for both samples with the passing current I = 60 mA in the Hall bar, the area of the domains magnetized up and down is approximately the same, which is associated with the action of the non-uniform Oersted field. At the same time, the found values of $M_r$ on the demagnetization curves (Fig. 3.) do not reach zero due to the non-uniform switching of the magnetization in the transverse part of the Hall bar. In Fig. 4 for the case I = +60 mA, it can be seen that there is a bright contrast in the lower transverse part of the Hall bar, and only a dark contrast in the upper one. This asymmetry in the transverse part of the Hall Bar leads to a small magnitude of $M_r$ not equal to 0.

Comparing the magnetization reversal curves for the samples with and without W, it can be noted that, despite the different values of the coercive force, the change in the magnetization from the saturation state occurs approximately at the same current values. This is due to the fact that the area of the Hall bar crossline, where the level of the AHE signal depends on the magnetization, begins to switch in the Ru/Co/Ru structure earlier.

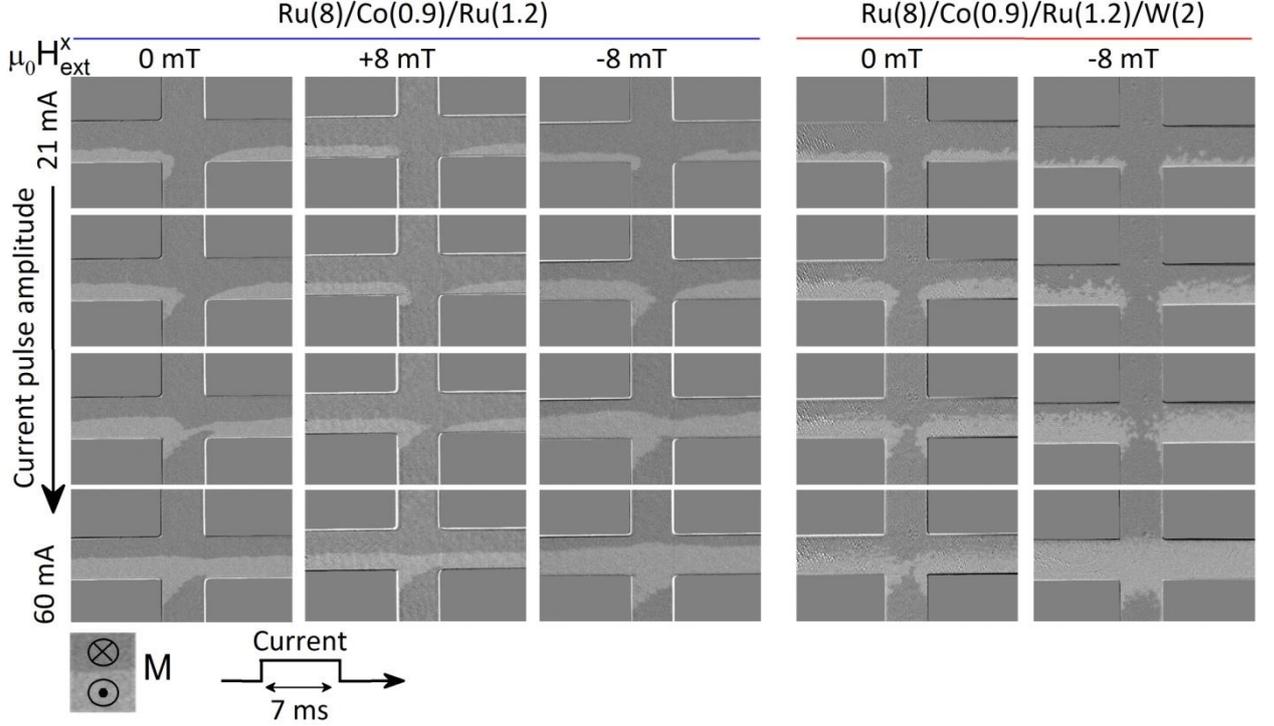

**FIG.4.** Changes in the domain structure during the current-induced magnetization reversal of Ru/Co/Ru and Ru/Co/Ru/W samples for different values of the field $\mu_0 H_{ext}^x$

2. *Current-induced magnetization reversal in an external field $\mu_0 H_{ext}^x = \pm 8 \, mT$*. As in the previous experiment, the sample was magnetized to the saturation state with an external field $\mu_0 H_{ext}^z$, after which it was turned off. However, now the current pulses were applied in the presence of an in-plane magnetic field oriented parallel to the current direction. In this case, the $M_r$ value after passing the current depends on its direction.

In the case of the Ru/Co/Ru structure, if the directions of the current and the field coincide, the region of the microstrip, in which the magnetization is switched, decreases to $M_r^+$ as compared with the case $\mu_0 H_{ext}^x = 0$, Fig.3(a). If the directions of the current and the field are opposite, the switching region is increased to $M_r^-$. Such a change is associated with the appearance of a nonzero effective field $\mu_0 H_{SOT}^z$, due to the SOT effect. Unlike the Oersted field, whose z-component on opposite sides of the microstrip has the opposite orientation, the SOT effect field has the same orientation over the entire area of the microstrip, due to the direction of magnetization tilt caused by action of the field $\mu_0 H_{ext}^x$ and by the direction of the spin current [30].

As seen in the Kerr microscopy images in Fig.4 for the Ru/Co/Ru/W structure, the current passing in the presence of the field $\mu_0 H_{ext}^x = -8 \, mT$ led to a complete switching of the magnetization (blue curve in Fig.3(b)). In the field $\mu_0 H_{ext}^x = +8 \, mT$ no change in the magnetization is observed (red line in Fig.3(b)). The illustration is not given for this case, since there is no change in the magnetic structure.

3. *Current-induced magnetization reversal in the presence of a constant field $\mu_0 H_{ext}^x$, but without preliminary magnetization by an external magnetic field to the saturation state*. By applying current pulses from -60 to +60 mA, we obtained current-induced hysteresis loops (marked by dashed lines in Fig. 3), in which the magnetization varies from $M_r^+$ to $M_r^-$. At the same time, the change in magnetization $\Delta M_r$ for the Ru/Co/Ru sample is 22%, and for the Ru/Co/Ru/W one is 80% of the change in magnetization during the complete magnetization reversal of the structure from the saturation state. However, to define that the current-induced magnetization reversal is incomplete, one has to visually observe the domain structure. In many previous studies based on measurement of AHE loops, this fact did not verified.

A change in the orientation of the field $\mu_0 H_{ext}^x$ leads to a symmetric reversal of the loop, which is characteristic of the magnetization reversal through the SHE mechanism [31]. For all the samples considered in the work, it was found that applying an external field $\mu_0 H_{ext}^y$ orthogonal to the current direction does not lead to the appearance of a current-induced field $\mu_0 H_{SOT}^z$. Based on this fact, it can be concluded that in the considered system only the longitudinal damping-like field (DL) [30] is responsible for switching. Therefore, when describing the effective fields of the SOT effect, instead of $\mu_0 H_{SOT}$, we will use the notation $\mu_0 H_L$.

Considering the details of the magnetization reversal process of the Ru/Co/Ru sample, it can be noted that the domain growth takes place asymmetrically, and the domain wall itself has a smooth profile as compared with the sample containing the W layer. We assume that these features are associated with the presence of the largest Dzyaloshinskii-Moriya interaction in the sample [17]. In our previous study, the value of DMI D= $0.2\ mJ/m^2$ was measured [32]. In the Ru/Co/Ru/W samples, these features disappear, which may be due to the absence of DMI.

The above study led to the following conclusions: (i) In the considered structures, the magnetization switching under the action of current always starts from the edge and, as the amplitude of the current increases, the domain wall moves parallel to the edge of the Hall bar; (ii) Measurement of AHE does not allow us to unambiguously judge the behavior of the magnetization and the degree of saturation of the sample; (iii) The sign of the effective field of the SOT effect is the same in both samples; (iv) Addition of the W layer in the considered structure enables a complete magnetization switching under the action of current.

### 3.2. Quantitative evaluation of current-induced magnetization reversal

For a quantitative description of the studied structures, an estimate of the effective magnetic fields generated by passing a current was made. The study showed that determining this value for low-coercive samples with a small value of the PMA energy is a nontrivial task. Below we consider three approaches to determine the $\mu_0 H_L$ value.

#### 3.2.1. Second harmonic generation measurement method

The harmonic Hall voltage measurement [33] is the generally accepted method for determining the SOT effect efficiency. During changing the magnitude of the external magnetic field $\mu_0 H_{ext}^x$, an alternating current is passed through a Hall bar and the transverse potential difference of AHE is recorded at the excitation frequency $V_\omega$ and at the double frequency $V_{2\omega}$. The analytical model [13], describing this process, allows to determine the magnitude of the effective magnetic field $\mu_0 H_L$, in the simplest case, by the formula:

$$\mu_0 H_L = -2\left(\frac{dV_{2\omega}}{\mu_0 dH_{ext}^x}\right)\bigg/\left(\frac{d^2 V_\omega}{\mu_0 dH_{ext}^{2x}}\right) \quad (1)$$

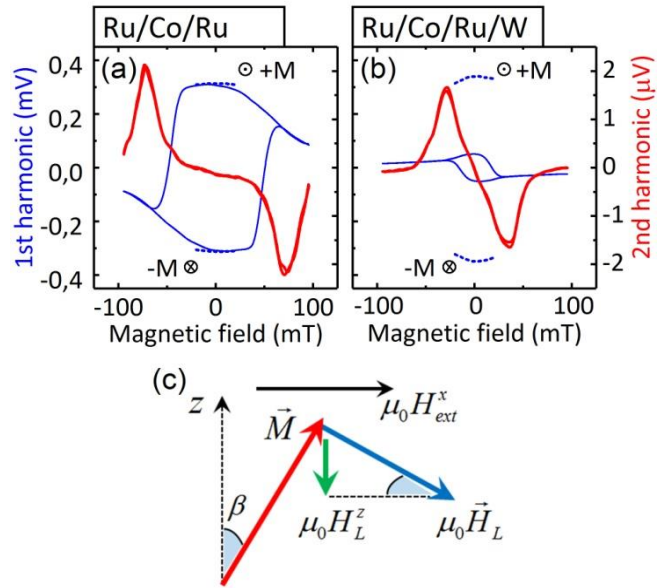

**FIG.5.** Dependence of the AHE amplitude on the magnitude of the in-plane field applied along the current direction at the first and second harmonics for structures without W (a) and with W (b). The signal at the first harmonic was also measured in a field range of ± 20 mT after the preliminary magnetization (dashed lines). (c) Scheme of mutual orientation of the magnetization, the effective field of SOT effect and its projection on z axis

Using this approach, experimental dependences of the AHE response on the first and second harmonics were obtained for the samples considered in this paper, Fig.5. During measurements, an alternating current with an amplitude of 10 mA and a frequency of 17 Hz was applied to the samples, and the external magnetic field changed along the current direction. The key parameter of these measurements is the slope of the linear segment of the second harmonic near zero field, since it corresponds to the magnitude of the magnetization deflection under the current action. It is seen that the addition of the W layer leads to a sharp increase in the slope, which indicates a significant increase in the effectiveness of the current action on the magnetization, Fig.5. For the numerical estimate of the effective field, in addition to the slope, the curvature of the first harmonic is determined. It was found that a change in the external field in the range of ± 100 mT leads to a partial demagnetization of the sample, which makes the resulting dependence incorrect. To study the response from the saturated state, the sample was preliminarily magnetized by an external perpendicular field, after which it was turned off, and the measurement was performed in the in-plane field range $\mu_0 H_{ext}^x = \pm 20\ mT$. As it can be seen in Fig.5, such an approach allows one to define very different values of the level and curvature of the first harmonic dependence (blue dashed lines). It can be assumed that the demagnetization of the samples is due to the small value of the effective field of PMA and a relatively large alternating current, while the combination of the external field and the induced field of the SOT effect was sufficient for demagnetization. For the case of the structure containing the W layer, where the SOT effect field is obviously larger, the demagnetization turned out to be stronger.

Using formula (1), we found the values of the effective field $\mu_0 H_L$ equal to 3 and 6 mT at the current of 60 mA for structures without and with W, respectively. The resulting field is oriented perpendicular to the magnetization, as shown in Fig. 5(c). Taking into account the magnitude of the anisotropy field (Table 1) and the conditions of the previously described experiment (Fig. 3), we can estimate the angle $\beta = 3.3^0$ и $7.6^0$ for the first and second structures, respectively. The projection of the $\mu_0 H_L$ field on the z-axis gives the corresponding values of $\mu_0 H_L^z$ equal to 0.13 and 0.8 mT. Comparing these values with magnetic hysteresis loops (Fig. 1(b)), it can be seen that there are not enough such fields for complete magnetization reversal as was observed in the sample containing the W layer. Thus, the second harmonic generation method does not allow to determine real values of the effective fields induced by a passing current.

### 3.2.2. Current-induced magnetization curve method

A qualitatively different approach for determining the effective field induced by the SOT effect is based on recording the magnetization curves [16]. The Ru/Co/Ru/W sample was demagnetized by an in-plane alternating field and then saturated in one case by means of an external perpendicular field Fig.6(a), and in another - by passing a current in the presence of a small planar field, Fig.6(b). Analyzing two magnetization curves, we can assume that the work on magnetization (Zeeman energy) in both cases should be the same:

$$E_z^H = M_s \int_0^1 \mu_0 H_{ext}^z dM_n \qquad (2)$$

$$E_z^I = M_s \alpha \int_0^1 I dM_n \qquad (3)$$

where $M_n$ is the normalized value of magnetization. Since the field induced by the current linearly depends on the current magnitude $I$, we can represent the effective field as $\mu_0 H_L^z = \alpha I$. Equating expressions (2) and (3), we can find the proportionality coefficient between the field and the current, which will be equal to the ratio of the areas $S_B$ and $S_I$ corresponding to the red and blue areas in Fig.6. The presented measurements were carried out only for the Ru/Co/Ru/W structure, since for the structure without W it is impossible to obtain a saturation state by means of current. As seen in Fig.6(b), at the initial stage the application of current does not lead to the exit from the saturation state. It can be assumed that this is due to the action of the non-uniform Oersted field, under the action of which the domains with different magnetization orientations are redistributed in the microstrip, while the resulting value of the magnetization remains zero. This process could not be illustrated using Kerr microscopy. However, it can be noted that at the initial stage the work on magnetization is not performed, so we did not take into account the area above the zero magnetization value. This approach allowed us to define the value $\mu_0 H_L^z = 4.4\ mT$, which, taking into account the mutual orientation of the fields, as in Fig. 5c, gives the value of the effective field $\mu_0 H_L = 33.8\ mT$.

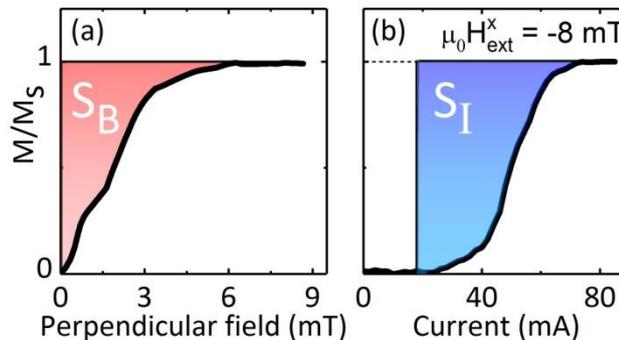

**FIG.6.** Magnetization curves of the Ru/Co/Ru/W structure measured under the action of the field (a) and current (b)

The action of such a field could lead to a complete magnetization switching; however, the considered method does not allow determining the value of the effective field in the sample without the W layer. For this purpose, we proposed an alternative approach.

### 3.2.3. Method of analyzing of the domain wall position

Analyzing the demagnetization process of Hall bars of both compositions under the current action in the absence of an external field (Fig.4), we can draw two conclusions: (i) the magnitude of the Oersted field generated by the current of the amplitude considered is greater than the magnitude of the coercive force of the both samples. This leads to the fact that the magnetization reversal occurs by moving the domain wall parallel to the long side of the Hall bar (perpendicular to the current); (ii) this movement under the action of the non-uniform Oersted field will continue to the point where the field value becomes zero, that is, to the middle of the Hall bar.

In the presence of an external field $\mu_0 H_{ext}^x$, the character of the magnetization reversal does not change: an increase in the current amplitude leads to parallel movement of the domain wall, but now, depending on the direction of the external field, the final position of the domain wall will be on opposite sides relative to the middle of the Hall bar. Such a change is associated with the appearance of an additional magnetic field $\pm\mu_0 H_L$, induced by the SOT effect, which is superimposed on the non-uniform Oersted field. It can be assumed that the movement of the domain wall will continue as before to the region with zero value of the resulting magnetic field, then the condition for the position of the domain wall can be written in the form:

$$\mu_0 H_{Oe} \pm \mu_0 H_L = 0 \tag{4}$$

The arguments presented correspond well to the behavior of the domain structure of the Ru/Co/Ru sample, Fig.4. The profile of the domain wall in the sample with W ceases to be smooth, which casts doubt on the applicability of the model described above, but the final conclusions can be applied to it.

Using condition (4), one can determine the SOT effect effective field, if one knows the position of the domain wall and the spatial distribution of the Oersted field. It can be noted that the magnitude and distribution of the Oersted field depends only on the geometry of the microstrip and the current strength. For further consideration, an analytical expression for the distribution of the Oersted field was obtained. Without taking into account the area of the Hall bar's crossline, we can assume that the current density is uniform over the cross section of the microstrip. Let us to consider a microstrip of length L and width b, Fig.7(a). We divide the microstrip into infinitely thin conductors with width $dy$ parallel to the current direction. Based on the Biot-Savart-Laplace law (5) (where $r$ is a radius-vector between a current element $dl$ and a point where the field is computed), one can obtain an expression for the magnetic field $\mu_0 H_{Oe}$ oriented perpendicular to the microstrip plane(6):

$$\mu_0 dH_{Oe} = \frac{\mu_0}{4\pi}\frac{I[\overline{dl}\times\vec{r}]}{r^3} \tag{5}$$

$$\mu_0 dH_{Oe} = \frac{\mu_0}{2\pi}\frac{dI}{y}\cos\alpha \tag{6}$$

Eq.(6) allows calculating the field value relative to the middle point of a conductor of length L with the passing current $dI$ in a point shifted on distance $y$. One can express the magnitude of the current passing through the elementary conductor through its width: $dI = (I/b)dy$. Each conductor contributes to the resulting magnetic field at point $y$:

$$\mu_0 dH_{Oe} = \frac{\mu_0 I}{2\pi b}\frac{1}{y}\cos\alpha\, dy = \frac{\mu_0 I}{2\pi b}\frac{1}{y}\frac{dy}{\sqrt{1+\tan^2\alpha}} = \frac{\mu_0 I}{2\pi b}\frac{1}{y}\frac{dy}{\sqrt{1+\frac{y^2}{(L/2)^2}}} \tag{7}$$

where, in accordance with the scheme of Fig.7(a), we moved from $\cos\alpha$ to distance parameters. Further, it is taken into account that the parts of the microstrip located on opposite sides of the point y, contribute with the opposite sign to the resulting magnetic field. Considering this when integrating one can get:

$$\mu_0 H_{Oe} = \frac{\mu_0 IL}{4\pi b}\left[\int_y^0 \frac{1}{y\sqrt{(L/2)^2+x^2}}dy - \int_{b-y}^0 \frac{1}{y\sqrt{(L/2)^2+y^2}}dy\right] = \frac{\mu_0 I}{2\pi b}\ln\left[\frac{(L/2)+\sqrt{(L/2)^2+(b-y)^2}}{(L/2)+\sqrt{(L/2)^2+y^2}}\frac{y}{b-y}\right] \tag{8}$$

This dependence describes the distribution of z-component of the Oersted field in a cross section of the microstrip perpendicular to the current direction. The calculation for different cases of length L showed that the condition L >> b is met for the experimentally considered Hall bars. With this in mind, the expression (8) is reduced to the form:

$$\mu_0 H_{Oe} = \frac{\mu_0 I}{2\pi b}\ln\left[\frac{b-y}{y}\right] \tag{9}$$

Analysis of expression (9) shows that when approaching the microstrip edges, the field value starts to increase asymptotically, so the maximum field depends on the area of homogeneity (step) along the *y*-axis. The calculation of the Oersted field distribution for two current values at a step of 50 nm is given in Fig.7(b). The obtained distribution qualitatively and quantitatively coincides with the distribution found by the simulation method as in [18]. To verify the compliance of the calculation results with the experiment, the demagnetization of the Hall bar at the initial stage was considered. The visualization of the change in the magnetic structure of the Ru/Co/Ru/W sample when exiting from the saturation state under the action of an external perpendicular field is shown in Fig.8. The structure with the W layer was considered due to the fact that the demagnetization process in it began from the edges, whereas in the Ru/Co/Ru sample from the middle. This case was compared to the case of demagnetization under the current action in the absence of an external field, that is, under the Oersted field action only. Roughly the effects of 1 mT and 20 mA are comparable. The distribution of the Oersted field for 20 mA is given in Fig.7(b), and at the edge of the microstrip it corresponds to a field of 1.2 mT. With such a comparison, an exact value cannot be found due to a number of factors: the non-uniformity of the Oersted field, the uncertainty of its magnitude at the edges, heating from the current, etc. However, for a rough estimate, we have tried to apply the analytically defined distribution.

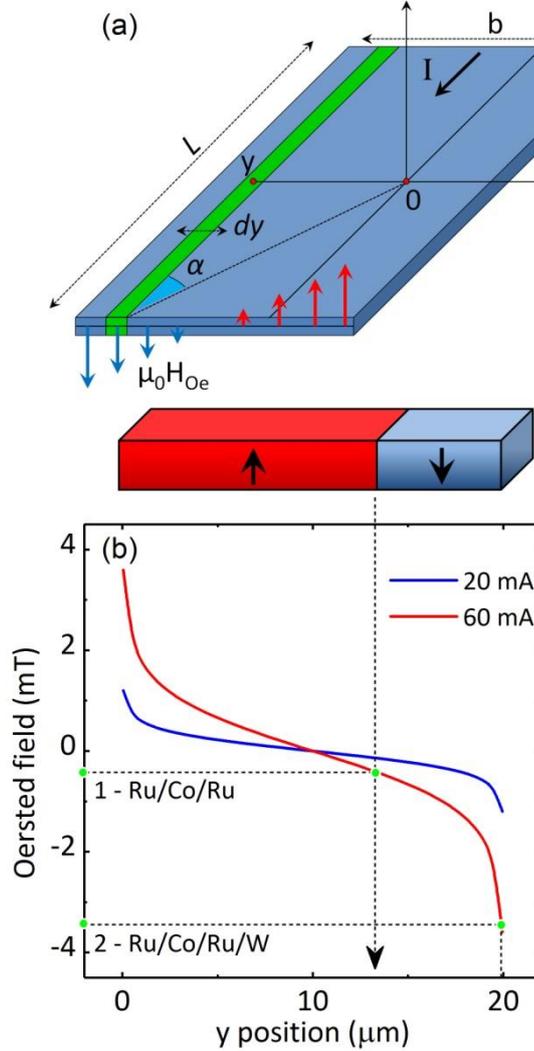

**FIG.7.** (a) Calculation scheme for the Oersted field on a homogeneous section of the Hall Bar. (b) The calculated distribution of the Oersted field in the cross section of the microstrip, perpendicular to the current direction. The vertical dashed lines mark the position of the domain wall for two structures after passing the maximum current into the presence of an external field $\mu_0 H_{ext}^x = 8\ mT$

Describing the position of the domain wall by condition (4), we can conclude that it will stop at the place of the Hall bar cross section, where z-component of the SOT field is modulo the Oersted field. Knowing the value of the Oersted field at each point of the Hall Bar and determining the position of the domain wall using Kerr microscopy (Fig.4), one can determine the field $\pm\mu_0 H_L$. For the sample without W, passing a current of 60 mA in the presence of a field of +8 and -8 mT gives the domain wall offset from the lower edge by 8.2 and 12.3 μm, respectively. The value of the Oersted field, and hence the resultant field of the SOT effect, in these places is -0.25 and +0.31 mT. In one case, the SOT effect field helps to exit from the saturation magnetization state, in the other - it prevents. The difference in values is due to the ambiguity of determining the position of the domain wall due to the nonlinearity of the boundary. The average value of z-component of the induced field is 0.28 mT, taking into account the magnetization tilting angle the magnitude of the effective field is 4.9 mT.

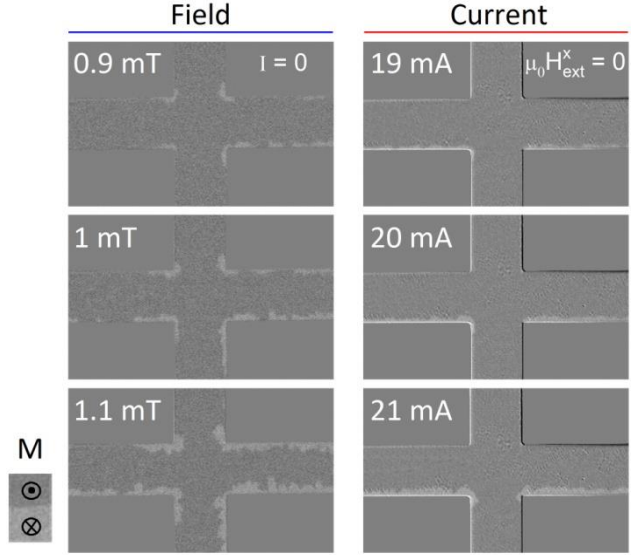

**FIG.8.** Kerr microscopy illustration of the initial stage of the Ru/Co/Ru/W Hall bar demagnetization under the action of the perpendicular field and current

As mentioned earlier, the external in-plane field of -8 mT is minimal for achieving a complete magnetization switching of the structure with W. With this field $\mu_0 H_{ext}^x$, the SOT effect field is everywhere larger than the Oersted field, i.e. $\mu_0 H_L^z \geq 3.6\ mT$. The corresponding value of the SOT effect effective field in the structure with W is 27.7 mT.

### 3.2.4. Comparison of results

The values of the effective fields found by different methods are given in Table 2. It can be seen that for the Ru/Co/Ru structure, the harmonic method and the method based on the distribution of the Oersted field are in good agreement. The results for the structure with the W layer are more scattered. For the final evaluation of the SOT effect in our structures, the values determined by the third method (domain wall position) were considered, since it is the most direct approach for the evaluation of the current-induced field.

**Table II**

The values of the effective fields $\mu_0 H_L$ (and their projections on the z axis - $\mu_0 H_L^z$) in mT, determined by three methods for samples of two compositions, and the SOT efficiency

| Sample | Methods of definition of $\mu_0 H_{SL}$ ($\mu_0 H_{SL}^z$) in mT | | | $\gamma_L$, $10^{-11}\ mT/Am^{-2}$ | $\xi_L$ |
|---|---|---|---|---|---|
| | Harmonic measurements | Magnetization curve | Domain wall analysis | | |
| Ru/Co/Ru | 3 (0.13) | - | 4.9 (0.28) | 1.6 | 0.03 |
| Ru/Co/Ru/W | 6 (0.8) | 33.8 (4.4) | 27.7 (3.6) | 9.2 | 0.15 |

For the considered systems, the coefficient showing the ratio of the magnitude of the induced field to the current γ and the SOT efficiency $\xi_L$ (effective angle of SHE) was calculated. The value of the SOT efficiency can be used for comparing the SOT effects in different structures [34]:

$$\xi_L = \frac{2e}{\hbar}\mu_0 M_s t_{Co}^{eff} \left|\frac{H_L}{j}\right| \quad (10)$$

where *e* and *ℏ* are the elementary charge and the Dirac constant. This value reflects the ratio of the spin current density adsorbed by a ferromagnetic layer to the charge current injected into a heavy metal layer. The values of current density *j* and effective field $\mu_0 H_L$ were substituted by values experimentally defined for *I*=60 mA. The effective thickness of the ferromagnetic layer was determined as the thickness of the Co layer equal to 0.5 nm taking into account the magnetic dead layer of 0.2 nm at each interface with Ru [32].

Assuming that the top Ru layer thickness is 1.2 nm and its contribution to the SOT effect is negligible, the efficiency of the Ru buffer layer is $\xi_L^{Ru} = 0.03$. Comparison of the defined value with previous works is difficult because very few papers are devoted to the study of the SOT effect in structures with Ru. We can point out the study of the epitaxial Ru/CoFeAl/MgO structure [6]. It is shown that with increasing thickness of the Ru buffer layer in the range up to 8 nm, the SOT effect effective fields increase reaching maximum values $H_L = 0{,}2 \cdot 10^{-11} mT/Am^{-2}$ and $H_T = 0{,}5 \cdot 10^{-11} mT/Am^{-2}$ with the maximum SOT efficiency $\xi_j = 0.0056$. The transverse (field-like, $H_T$) field turned out to be larger than the longitudinal (antidamping-like, $H_L$), whereas in our work it was not detected at all, while the value of the longitudinal field in the Ru/Co/Ru structure is almost an order of magnitude larger.

Taking into account the measured value of $\xi_L^{Ru}$, the SOT efficiency of the W layer in the Ru/Co/Ru/W structure is $\xi_L^W = 0.12$. The presence of an intermediate Ru/Co interface leads to additional screening due to the high spin-memory-loss parameter $\delta_{Co/Ru} = 0.34$ [35, 36], so the value of $\xi_L^W$ is slightly less than the previously defined data for the mixed α + β crystalline phase of W: $\xi_L^W = 0.18$ [20].

## IV. CONCLUSIONS

An experimental study of the current-induced magnetization reversal of the Ru/Co/Ru and Ru/Co/Ru/W structures have been performed. In both structures, only a longitudinal field (antidumping-like) was generated, the ratio of the effective field to the current and the SOT efficiency are $1.6 \cdot 10^{-11} mT/Am^{-2}$ and 0.03 for the sample without W and $9.2 \cdot 10^{-11} mT/Am^{-2}$ and 0.15 for the sample with W. It is shown that the determination of the magnitude of the SOT effect effective fields in samples with small values of the coercive force and the anisotropy field is a nontrivial problem. Observation of the domain structure during the magnetization reversal process allows to explain the features on the AHE loops. It is impossible to accurately judge the magnetization behavior inside a Hall bar and the degree of magnetization saturation of the sample from the AHE loops. A direct method for estimating the magnitude of the SOT effect field is proposed based on the analysis of the domain wall position in the case of its displacement parallel to the current combined with the spatial distribution of the Oersted field. It is shown that for the implementation of the SOT effect one has to use an additional 5d metal layer with a strong SHE, while the rest of the structure, based on 4d metals, determines the magnetic parameters of the ferromagnetic layer. The advanced method for estimation of the SOT effect efficiency in low coercivity ferromagnetic multilayers can be adopted to a wide spectrum of materials.

## ACKNOWLEDGMENTS

This work was supported in part by Grant Council of the President of the Russian Federation (Grant No. MK-3650.2018.9), by Russian Foundation for Basic Research (Grants No. 18-32-00867, No. 18-52-53038), by the Russian Ministry of Education and Science under the state task (Grants No. 3.5178.2017/8.9 and No. 3.4956.2017), by Act 211 of the Government of the Russian Federation (Contract No. 02.A03.21.0011).